\documentclass{Interspeech}
\usepackage{amssymb}
\usepackage{multirow}
\usepackage{colortbl}
\usepackage{xcolor}
\usepackage{tabularx}
\usepackage{hyperref}

\interspeechcameraready

\title{Neural Codec Source Tracing: Toward Comprehensive Attribution in Open-Set Condition}

\author[affiliation={1}\dagger]{Yuankun}{Xie}
\author[affiliation={2,3}\dagger]{Xiaopeng}{Wang}
\author[affiliation={2,3}]{Zhiyong}{Wang}
\author[affiliation={3}]{Ruibo}{Fu}
\author[affiliation={4}]{Zhengqi}{Wen}
\author[affiliation={5}]{Songjun}{Cao}
\author[affiliation={5}]{Long}{Ma}	
\author[affiliation={6}]{Chenxing}{Li}	
\author[affiliation={1}]{Haonnan}{Cheng}
\author[affiliation={1}]{Long}{Ye}

\affiliation {State Key Laboratory of Media Convergence and Communication}{Communication University of China}{China}
\affiliation{School of Artificial Intelligence}{University of Chinese Academy of Sciences}{China}
\affiliation{Institute of Automation}{Chinese Academy of Sciences}{China}
\affiliation{Beijing National Research Center for Information Science and Technology}{Tsinghua University}{China}
\affiliation{YouTu Lab}{Tencent}{China}
\affiliation{AI Lab}{Tencent}{China}

\address{
	\small $^1$, Communication University of China
	\small $^2$ Institute of Automation, Chinese Academy of Sciences 
	\small $^3$ School of Artificial Intelligence, University of Chinese Academy of Sciences
	\small $^4$ Department of Automation, Tsinghua University
	\small $^5$ Beijing National Research Center for Information Science and Technology, Tsinghua University}
\email{xieyuankun@cuc.edu.cn}

\keywords{audio deepfake detection, neural codec, audio deepfake source tracing, out-of-distribution detection}

\begin{document}
	
\maketitle
\renewcommand{\thefootnote}{} 
\footnotetext[1]{$\dagger$ denotes equal contribution to this work. }
\renewcommand{\thefootnote}{\arabic{footnote}}
\setcounter{footnote}{0}
\begin{abstract}

Current research in audio deepfake detection is gradually transitioning from binary classification to multi-class tasks, referred as audio deepfake source tracing task. However, existing studies on source tracing consider only closed-set scenarios and have not considered the challenges posed by open-set conditions. In this paper, we define the Neural Codec Source Tracing (NCST) task, which is capable of performing open-set neural codec classification and interpretable ALM detection. Specifically, we constructed the ST-Codecfake dataset for the NCST task, which includes bilingual audio samples generated by 11 state-of-the-art neural codec methods and ALM-based out-of-distribution (OOD) test samples. Furthermore, we establish a comprehensive source tracing benchmark to assess NCST models in open-set conditions. The experimental results reveal that although the NCST models perform well in in-distribution (ID) classification and OOD detection, they lack robustness in classifying unseen real audio. The ST-codecfake dataset\footnote{https://zenodo.org/records/14631091} and  code\footnote{https://github.com/xieyuankun/ST-Codecfake} are available.
\end{abstract}

\section{Introduction}
In recent years, there has been rapid advancement in the field of text-to-speech (TTS) and voice conversion (VC), which called deepfake audio. Diverse endeavors and competitions, such as ASVspoof \cite{liu2023asvspoof, wang2024asvspoof5} and Audio Deepfake Detection (ADD) challenge \cite{yi2022add, yi2023add}, have been instituted to promote research aimed at developing deepfake countermeasure (CM) \cite{xie23c_interspeech, wang2024genuine, wang2024generalized}. 

With the advancements in binary classification (real/fake) performance in the field of ADD, research has gradually shifted towards audio deepfake source tracing task \cite{ xie2024generalized, klein2024source}. Notably, the well-known source tracing competition, Audio Deepfake Detection Challenge 2023 Track 3 (ADD2023T3), has clearly defined the source tracing task, which involves not only identifying real and in-distribution (ID) fake audio but also considering an out-of-distribution (OOD) novel fake method as OOD. However, the task design and evaluation methods for source tracing models require further optimization.

Firstly, the ID real category does not consider the unseen real audio. Current evaluations of source tracing model overlook the ability of identifying unseen real audio, which is especially critical when OOD categories are characterized as unseen fake audio. For a well-performing source tracing model, it must exhibit high confidence in correctly classifying unseen real audio as ID real. Secondly, for the fake categoary, current approaches only consider a single novel fake as the OOD fake and lack a discussion on how deepfakes generated by different configurations, sources, and methods impact the performance of source tracing systems. Lastly, regarding the evaluation of source tracing models, the currently used threshold-based metric, F1-score, is advantageous for decision-making. However, the selection of the threshold is overly subjective, which can lead to biased evaluations of OOD methods.

As the deepfake audio type, most of the aforementioned ADD studies focus on vocoder-based audio. With the development of large language models, the majority of fake audio on social media today consists of audio language model (ALM)-based deepfake audio, which is typically synthesized using neural codec models as the back-end. As CMs, some researchers have employed neural codec models to reconstruct audio for effectively detecting ALM-based audio \cite{xie2024codecfake,wu2024codecfake}. However, these studies have primarily focused on binary classification for the real and codec-based audio and have not yet explored source tracing for codec-based audio.

\begin{figure}[!t]
	\centering
	\includegraphics[width= 2.5in]{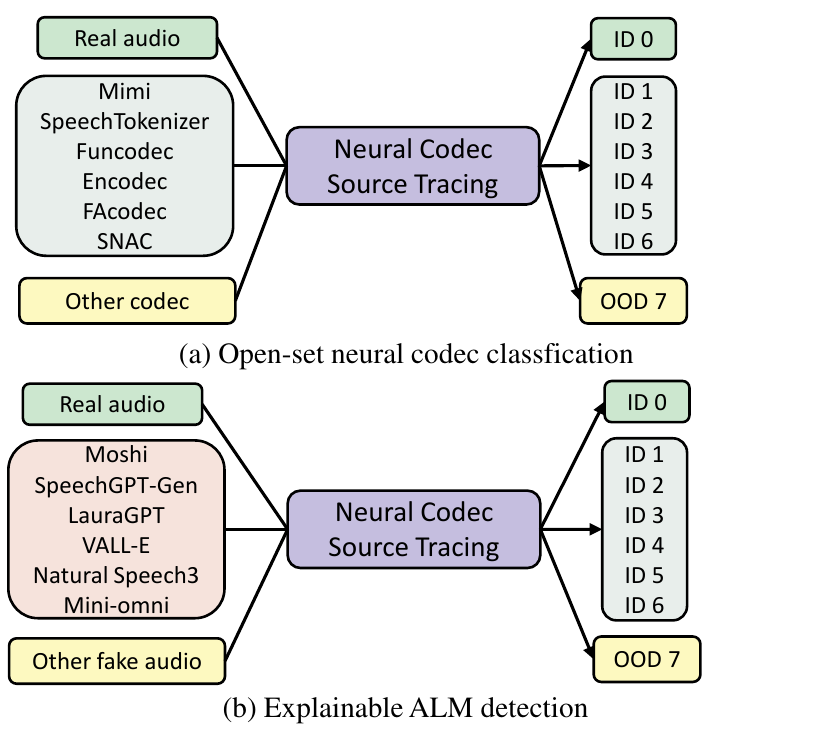}
	\hfil
	\caption{Two abilities of the neural codec source tracing method: (a) Open-set neural codec classification, (b) Explainable ALM-based audio detection.}
	\label{fig:ability}
	\vspace{-0.5cm}
\end{figure}

\begin{figure*}[!t]
	\centering
	\includegraphics[width= 6in]{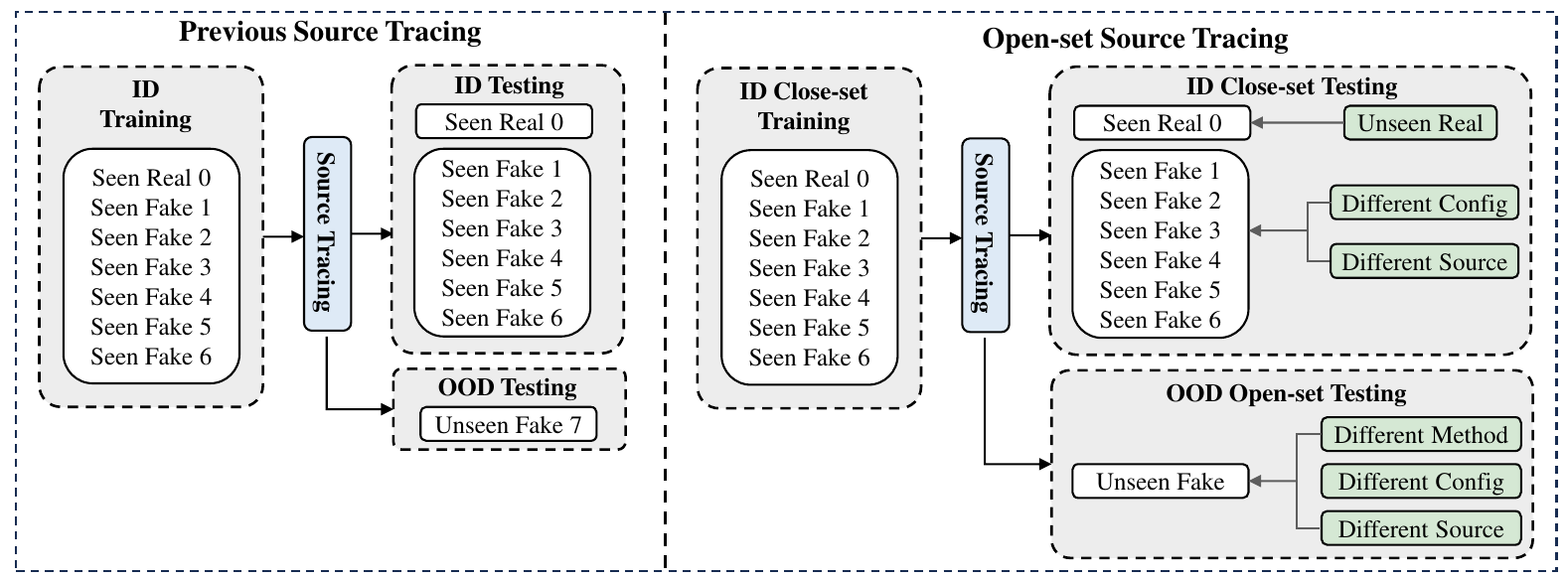}
	\hfil
	\caption{Open-set audio deepfake source tracing benchmark.}
	\label{fig:pipeline}
\end{figure*}

Source tracing for neural codecs involves analyzing the homogeneity (real vs. all neural codecs) and heterogeneity (within neural codec classes), as illustrated in Figure \ref{fig:ability}(a). Using a pre-trained neural codec source tracing (NCST) model, we can distinguish between real audio and different ID codecs, as well as identify OOD novel codecs. This capability can significantly enhance the protection of intellectual property for ID neural codec methods. Furthermore, as shown in Figure \ref{fig:ability}(b), a pre-trained NCST model can enhance the interpretability of ALM-based audio detection. For example, for recently developed ALM-based spoken dialogue models such as Moshi \cite{defossez2024moshi} and Mini-Omni \cite{xie2024mini}, the NCST can not only identify them as fake but also provide an explanation for the detection (e.g., their backend uses the Mimi \cite{defossez2024moshi} and SNAC \cite{siuzdak2024snac}).

In this paper, we focus on NCST task and further refine the evaluation conditions for open-set source tracing. First, we construct a dataset named \textbf{ST-Codecfake}, which includes 235,736 bilingual audio samples and wild ALM-based OOD test audio samples. Specifically, the real domain in ST-Codecfake consists of subsets from VCTK \cite{Veaux2017CSTRVC} and AISHELL3 \cite{shi21c_interspeech}, while the fake speech samples are generated by 11 different neural codec methods. Second, we establish an open-set source tracing benchmark, which evaluates source tracing models under both conventional ID and OOD classification conditions, as well as NCST model robustness evaluation. This includes analyzing the impact of unseen real audio on the ID real category, and the effects of fake audio generated using different methods, configurations, and sources on NCST model. For evaluation metrics, we adopt the F1-score for decision-making in ID testing. To measure OOD performance, we employ three threshold-free metrics to better evaluate the performance of source tracing systems on OOD data. 
Experiment results show that the best NCST model performs well in both ID classification and OOD detection, achieving an F1-score of 99.99\% for ID classification, an AUC of 97.54\% for open-set OOD detection, and an F1-score of 92.36\% for back-end detection of ALM-based audio.

\begin{table}[t]
	\caption{The details of ST-Codecfake dataset.}
	\label{tab:ST-Codecfake}
	\centering
	\begin{tabular}{llccc}
		\toprule
		\textbf{Type} &\textbf{Label} &\textbf{Train} & \textbf{Dev} & \textbf{Eval}  \\
		\midrule
		ID Real &0 &10000  & 1000  &13228 \\
		ID Fake &1-6 &60000  &6000  &79368 \\
		OOD Fake &7-11 &- &- &66140 \\
		Total &0-11 &70000 &7000 &158736 \\
		\bottomrule
	\end{tabular}
\end{table}

\section{Source tracing benchmark}
\label{benchmark}
Previous source tracing models, as shown on the left side of Figure \ref{fig:pipeline}, were trained on real audio and seen fake audio for ID classification. During testing, in addition to ID classification, one unseen fake audio was evaluate for OOD detection experiments. To comprehensively evaluate the performance of an NCST model, we first proposed ST-Codecfake dataset and established a new testing protocol for thorough assessment of NCST models.

\subsection{ST-Codecfake dataset}
To study the NCST task, it is essential to have relevant datasets. Considering the continuous updates in neural codec methods and ALMs, we propose a new dataset called ST-Codecfake. The real domain of ST-Codecfake is derived from subsets of VCTK and ALSHELL, including 10,000 non-overlapping training samples, 1,000 development samples, and 13,228 test samples. Utilizing these as the source domain, we constructed the ST-Codecfake dataset synthesis through 11 neural codec methods. Among these, we designated the commonly used neural codec in ALM as ID, which facilitates subsequent ALM-based audio detection tasks for interpretability. A detailed description of the dataset is provided in Table \ref{tab:ST-Codecfake}.

\subsection{ID close-set evaluation}
At this stage, we first train the NCST model on the ST-Codecfake training set (0-6) for 7-class classification. During the testing phase, we evaluate the model on the ST-Codecfake test set (0-6) for ID closed-set classification using the F1-score. Additionally, we conduct three extra experiments to ensure the robustness of the NCST model in ID classification:

\textbf{Unseen Real}: We test the robustness of the NCST model using unseen real audio. Specifically, we measure performance using the F1-score on the real sets from ASVspoof2019LA (19LA) \cite{nautsch2021asvspoof}, In the Wild (ITW) \cite{muller22_interspeech}, and a subset of NCSSD \cite{liu2024generative}. The 19LA real set contains 7,355 audio samples, sourced from the VCTK dataset, which features relatively clean audio. The ITW dataset includes 19,963 real audio samples from YouTube, representing a complex acoustic environment. For NCSSD, we select 10,000 real audio samples recorded in a studio environment, representing real audio from spoken dialogue scenarios.

\textbf{Different Configuration}: For seen fake audio, different generation configurations—such as sample rate, bitrate, and the number of quantizers—can lead to variations in the reconstructed audio. To investigate whether these variations impact the NCST system, we selected configurations C3-1 to C3-4 and C4-1 to C4-3 from Codecfake dataset \cite{xie2024codecfake}, which represent the different configurations of ID 3 FunCodec and ID 4 EnCodec.

\textbf{Different Source}: To investigate whether the seen fake methods from different source domains affect the NCST system, we used NCSSD as the real source domain and employed ID 1 Mimi and ID 6 SNAC for reconstruction.

\subsection{OOD open-set evaluation}
At this stage, in order to objectively evaluate the performance of NCST on OOD detection, we use three threshold-independent metrics: AUC (Area Under the Curve), FPR95 (False Positive Rate at 95\% True Positive Rate), and EER (Equal Error Rate). During the calculation of these metrics, the labels for OOD classes are set to 0, while the labels for ID classes (0-6) are set to 1. In this process, we consider the following three scenarios to comprehensively assess the OOD detection capability of the NCST model:

\textbf{Different Method}:
This section includes two test items: multi codec (7-11) and other fake type (19LA+ITW). For these two audio test samples, the NCST model needs to provide OOD results. The multi codec test is designed to explore whether the mixture of multiple OOD codec-based audio samples affects the OOD performance of the NCST model. For the other fake type, we selected non-neural codec synthesis methods, including 63,882 fake audio samples from the 19LA test set and 11,816 fake audio samples from ITW.

\textbf{Different Configuration}: Similar to the ID close-set evaluation, we choose to re-synthesize OOD 7 Wavtokenizer using different parameters to verify the robustness of the OOD detection capability. The original version of OOD 7 Wavtokenizer operates at 75 tokens/s, while the re-synthesized version uses 40 tokens/s.

\textbf{Different Source}: Similar to the ID close-set evaluation, we use NCSSD as the source domain and perform reconstruction using Wavtokenizer to verify the robustness of the NCST model's OOD detection capability.

\begin{table}[t]
	\caption{The details of Neural codec models of ST-Codecfake. B represents bandwidth and Q represents quantizers.}
	\label{tab:encodec}
	\centering
	\begin{tabular}{llccc}
		\toprule
		\textbf{Type} & \textbf{Neural Codec} & \textbf{SR} & \textbf{B} & \textbf{Q} \\
		\midrule
		ID 1  & Mimi \cite{defossez2024moshi}            & 24k & 1.1k  & 8  \\
		ID 2  & SpeechTokenizer \cite{zhang2024speechtokenizer}  & 16k & 4k    & 8  \\
		ID 3  & FunCodec \cite{du2023funcodec}          & 16k & 16k   & 32 \\
		ID 4  & EnCodec \cite{defossez2022high}          & 24k & 6k    & 8  \\
		ID 5  & FACodec \cite{ju2024naturalspeech}       & 24k & 6.4k  & 8  \\
		ID 6  & SNAC \cite{siuzdak2024snac}              & 24k & 0.98k & 3  \\
		OOD 7 & WavTokenizer \cite{ji2024wavtokenizer}   & 24k & 0.9k  & 1  \\
		OOD 8 & AcademicCodec \cite{yang2023hifi}        & 24k & 3k    & 4  \\
		OOD 9 & DAC \cite{kumar2024high}                 & 44k & 8k    & 9  \\
		OOD 10 & AudioDec \cite{wu2023audiodec}          & 24k & 6.4k  & 8  \\
		OOD 11 & SoundStream \cite{zeghidour2021soundstream} & 16k & 4k & 8  \\
		\bottomrule
	\end{tabular}
\end{table}

\subsection{Explainable ALM detection}
In the final stage, we tested the NCST model using ALM-based audio to determine whether the model can accurately classify ALM-based audio backends. This experiment includes four ALM methods: Moshi, Mini-omni, SpeechGPT-Gen \cite{zhang2024speechgpt}, and VALL-E \cite{wang2023neural}. Moshi and Mini-omni are proposed for spoken dialogue scenarios. We use 10,000 real audio samples from NCSSD as instructions and generated 10,000 deepfake audio samples for both Moshi and Mini-omni. For SpeechGPT-Gen, we obtained 18 audio samples of generated responses from the sample page. For VALL-E, we obtained 4,436 fake audio samples from Codecfake A1.

\section{Neural codec source tracing method}
\subsection{Baseline model}
This section introduces the baseline models used for the NCST task, including three representative models in the field of ADD: Mel-LCNN, AASIST, and W2V2-AASIST. We first consider the impact of different features in the NCST scenario. Therefore, we selected handcrafted feature Mel-Spectrogram with LCNN \cite{lavrentyeva2019stc} back-end, raw audio waveforms with AASIST \cite{jung2022aasist}, and frozen self-supervised features Wav2Vec-XLS-R\footnote{https://huggingface.co/facebook/wav2vec2-xls-r-300m} with AASIST (W2V2-AASIST). Notably, W2V2-AASIST achieved the best performance in the ADD2023 T3 single system in previous work \cite{xie2024generalized}. 

\subsection{OOD detection method}
We selected five state-of-the-art OOD methods for performance comparison, including MSP \cite{hendrycks2016baseline}, Energy \cite{liu2020energy}, KNN \cite{sun2022out}, Mahalanobis \cite{lee2018simple}, and NSD \cite{xie2024generalized}. Among these, MSP and Energy are OOD detection methods based on logits scores, while KNN and Mahalanobis are feature distance-based OOD detectors. NSD utilizes both features and logits for OOD decision-making.
\subsection{Implementation detail}
For the training and testing data, all audio samples were downsampled to 16,000 Hz and trimmed or padded to a duration of 4 seconds (64,600 samples). Consequently, we obtained Mel-Spectrograms of size (80, 404), raw audio waveforms of 64,600 samples, and W2V2 features of size (1024, 201) as the input features for the three NCST models. All baseline NCST models were configured with a classification head of ID 7 classes and trained using cross-entropy loss. We trained all the models for 50 epochs. The learning rate was initialized at $5^{-5}$ and halved every 10 epochs. The model that exhibited the best performance on the development set was selected as the evaluation model.

\begin{table}[t]
	\caption{F1-score (\%) results for ID classification experiment.}
	\label{tab:id}
	\centering
	\fontsize{7.5pt}{10pt}\selectfont
	\begin{tabular}{l*{8}{p{0.3cm}}}
		\toprule
		\textbf{NCST model} & \textbf{0} & \textbf{1} & \textbf{2} & \textbf{3} & \textbf{4} & \textbf{5} & \textbf{6} & \textbf{AVG} \\
		\midrule
		Mel-LCNN & 90.33 & 99.86 & 98.84 & 99.34 & 97.50 & 98.32 & 91.29 & 96.50 \\
		AASIST & 99.98 & 99.99 & 100 & 100 & 100 & 100 & 99.98 & 99.99 \\
		W2V2-AASIST & 99.99 & 100 & 100 & 100 & 100 & 99.99 & 99.98 & 99.99 \\
		\bottomrule
	\end{tabular}
\end{table}

\begin{table*}[t]
	\centering
	\caption{F1-score (\%) results for ID close-set robustness evaluation.}
	\label{tab:addlabel}
	\fontsize{8pt}{10pt}\selectfont
	\begin{tabular}{p{7.5em}ccccccccccccc}
		\toprule
		\multirow{2}[4]{*}{NCST model} & \multicolumn{3}{c}{Unseen real} & \multicolumn{7}{c}{Different config} & \multicolumn{2}{c}{Different source} & \multirow{2}[4]{*}{AVG} \\
		\cmidrule(lr){2-4} \cmidrule(lr){5-11} \cmidrule(lr){12-13}
		& 19LA & ITW & NCSSD & C3-1 & C3-2 & C3-3 & C3-4 & C4-1 & C4-2 & C4-3 & Mimi & SNAC & \\
		\midrule
		Mel-LCNN & 1.96 & 26.80 & \bf 21.67 & 98.65 & 99.03 & 98.98 & 98.98 & 96.05 & 66.25 & 8.07 & 93.33 & 51.12 & 63.41 \\
		AASIST & 0.73 & 24.83 & 8.00 & \bf99.99 & \bf100 & \bf100 & \bf100 & \bf99.98 & \bf99.95 \bf& \bf99.87 & \bf96.64 & \bf98.93 & 77.41 \\
		W2V2-AASIST & \bf 16.94 & \bf 77.01 & 4.77 & 98.99 & 99.7 & 99.8 & 99.8 & 99.96 & 99.71 & 98.19 & 84.08 & 83.65 & \bf 80.22 \\
		\bottomrule
	\end{tabular}%
\end{table*}

\begin{table*}[htbp]
	\centering
	\caption{AUC (\%) ↑, FPR95 (\%) ↓, EER (\%) ↓ results for OOD open-set evaluation.}
	\label{tab:ood}
	\fontsize{7.5pt}{10pt}\selectfont
	\begin{tabular}{p{1cm}*{16}{p{0.35cm}p{0.4cm}p{0.4cm}}}
		\toprule
		\multirow{2}[4]{*}{OOD} & \multicolumn{3}{c}{OOD single codec 7} & \multicolumn{3}{c}{OOD multi codec 8-11} & \multicolumn{3}{c}{OOD  (19LA+ITW)} & \multicolumn{3}{c}{Different config} & \multicolumn{3}{c}{Different source} & \multicolumn{3}{c}{OOD method AVG} \\
		\cmidrule(lr){2-4} \cmidrule(lr){5-7} \cmidrule(lr){8-10} \cmidrule(lr){11-13} \cmidrule(lr){14-16} \cmidrule(lr){17-19}
		& AUC & FPR95 & EER & AUC & FPR95 & EER & AUC & FPR95 & EER & AUC & FPR95 & EER & AUC & FPR95 & EER & AUC & FPR95 & EER \\
		\midrule
		MSP & \bf 99.93 & \textbf{0.17} & \textbf{0.90} & \bf 93.34 & \bf 22.40 & \bf 14.64 & \textbf{98.48} & \textbf{6.72} & \textbf{6.13} & 96.23 & 18.01 & 10.65 & 99.71 & 1.31 & 2.40 & \bf 97.54 & \bf 9.72 & \bf 6.94 \\
		Energy & \bf 99.93 & \textbf{0.17} & 0.97 & 92.83 & 23.29 & 15.29 & 98.33 & 7.14 & 6.42 & 96.46 & 17.06 & 10.24 & 99.74 & 1.15 & 2.33 & 97.46 & 9.76 & 7.05 \\
		KNN & 98.80 & 5.89 & 5.46 & 92.61 & 25.38 & 15.93 & 92.64 & 17.21 & 13.41 & 91.92 & 35.15 & 17.55 & 98.94 & 5.10 & 5.08 & 94.98 & 17.75 & 11.49 \\
		Mahalanobis & 94.34 & 16.81 & 11.62 & 90.69 & 28.21 & 17.83 & 93.57 & 20.08 & 14.29 & 76.00 & 48.49 & 30.74 & 94.83 & 16.09 & 11.47 & 89.89 & 25.94 & 17.99 \\
		NSD & 99.09 & 3.55 & 3.95 & 85.47 & 26.85 & 22.95 & 94.06 & 17.98 & 12.89 & \textbf{98.55} & \textbf{7.79} & \textbf{6.26} & \textbf{99.77} & \bf 0.65 & \bf 2.07 & 95.39 & 11.36 & 9.62 \\
		\midrule
		AVG & 98.82 & 5.32 & 4.58 & 90.99 & 25.23 & 17.33 & 95.82 & 13.03 & 10.23 & 91.83 & 25.90 & 15.89 & 98.99 & 4.86 & 4.67 & 95.65 & 14.51 & 10.62 \\
		\bottomrule
	\end{tabular}
\end{table*}

\begin{table}[t]
	\caption{F1-score (\%) results for ALM back-end detection.}
	\label{tab:alm}
	\centering
	\fontsize{6.8pt}{10pt}\selectfont
	\begin{tabular}{p{1.6cm} >{\centering\arraybackslash}p{0.4cm} >{\centering\arraybackslash}p{1.1cm} >{\centering\arraybackslash}p{0.9cm} >{\centering\arraybackslash}p{0.9cm} >{\centering\arraybackslash}p{0.35cm}}
		\toprule
		\textbf{NCST} & \textbf{Moshi} & \textbf{Mini-omni} & \textbf{SpeechGPT} & \textbf{VALL-E} & \textbf{AVG} \\
		\midrule
		Mel-LCNN   & \bf 97.70 & \bf100.00 & 68.03 & 99.88 & 91.40 \\
		AASIST     & 10.98 & \bf 100.00 & \bf 99.48 & \bf100.00 & 77.62 \\
		W2V2-AASIST & 70.74 & \bf 100.00 & 99.35 & 99.35 & \bf 92.36 \\
		\bottomrule
	\end{tabular}
\end{table}

\section{Results and Discussion}
\subsection{ID close-set evaluation}
After training with ST-codecfake, we conducted evaluations as described in Section \ref{benchmark}. First, we evaluated the NCST model on the ST-Codecfake test set (0-6) for ID closed-set classification using the F1-score. The results are shown in Table \ref{tab:id}. It can be observed that in the ID classification experiments, all three baseline models achieved promising results, with AASIST and W2V2-AASIST reaching an average F1-score of 99.99\%.
 
To further validate the robustness ID performance of the NCST models, we conducted three additional experiments: Unseen Real, Different Configuration, and Different Source experiments. Among the models, W2V2-AASIST achieved the best average F1-score of 80.22\%. It is noteworthy that all NCST models performed poorly on the Unseen real test, with even the best model, W2V2-AASIST, achieving only 16.94\%, 77.01\%, and 4.77\% F1-scores on the 19LA real subset, ITW real subset, and NCSSD set, respectively. We attribute the bad performance to two main factors: the insufficient amount of real training data and the potential underrepresentation of real class weights. In the Different Configuration and Different Source experiments, AASIST and W2V2-AASIST achieved favorable results for ID neural codecs, except for Mel-LCNN, which showed significant performance declines in the C4-3 and SNAC.

\subsection{OOD open-set evaluation}
At this stage, we conducted OOD open-set evaluation on the W2V2-AASIST NCST model, which had the best performance in the ID evaluation phase. We employed five different OOD methods for this evaluation. The results are shown in Table \ref{tab:ood}. For OOD method average score (calculated horizontally in Table \ref{tab:ood}), the logits-based methods, MSP and Energy, outperformed the feature-based and logits-based NSD, which in turn performed better than the feature-based KNN and Mahalanobis methods. Among these, MSP achieved the best overall performance across all conditions, with an AUC of 97.54\%, an FPR95 of 9.72\%, and an EER of 6.94\%. This indicates that logits-based OOD method are sufficient to distinguish between ID and OOD novel codecs. By observing the logits of test samples, we found that the NCST model produces excessively high-confidence logits for ID samples. This results in low-confidence logits for both unseen real and unseen fake samples, making them easily distinguishable. Consequently, this leads to strong OOD performance but a decrease in performance for unseen real data.

As the average scores of all OOD methods for each OOD test condition (calculated vertically in Table \ref{tab:ood}), we found that the AUC for each OOD evaluation condition exceeded 90\%. Specifically, the simplest case was a single method considered as the OOD, namely OOD single codec 7 condition, which achieved the best performance with an AUC of 98.82\%. Under the different configuration and different source conditions of neural codec 7, the W2V2-AASIST NCST model achieved AUC of 91.83\% and 98.99\%, respectively.
Extending the scenario where a single codec is considered as OOD, we also evaluated NCST model where multiple neural codecs were mixed and treated as OOD. This test produced an AUC of 90.99\%, representing the most challenging scenario and suggesting that detecting a diverse mix of novel codecs is more difficult. Additionally, we examined whether the NCST model could identify 19LA fake and ITW fake as OOD, with an average AUC of 95.82\%, indicating that these cases are relatively easy to distinguish.

\subsection{Explainable ALM detection}
Finally, we conducted experiments to evaluate the NCST model's accuracy in detecting the back-end of ALM. The results, shown in Table \ref{tab:alm}, demonstrate that the Mel-LCNN, AASIST, and W2V2-AASIST models achieved F1-scores of 91.40\%, 77.62\%, and 92.36\%, respectively. Among these, Mini-omni was the easiest to detect, achieving a perfect F1-score of 100\% in 10,000 test samples. 

\section{Conclusion}
In this paper, we focus on the NCST task and establish a comprehensive benchmark for evaluating source tracing models. To this end, we first constructed the ST-codecfake dataset and used it to train three baseline NCST models, which were then tested in ID close-set and OOD open-set conditions. The experimental results demonstrate that although the NCST models perform well in ID classification and OOD detection, they lack robustness in real classification, often misclassifying unseen real speech. Future work will aim to optimize the models for improved robustness in real category classification.

\tiny
\bibliographystyle{IEEEtran}
\bibliography{mybib}

\end{document}